\documentclass[10pt, aas2pp4, flushrt, mathptmx]{aastex}

\usepackage{epsf}

\newcommand{\ha}{H$\alpha$~}

\begin{document}

\sffamily

\title{On the origin of intergranular jets}

\author{ V. B. Yurchyshyn\altaffilmark{1}, P.R. Goode\altaffilmark{1}, V. I.
Abramenko\altaffilmark{1} \& O. Steiner\altaffilmark{2}}

\affil{$^1$ \it Big Bear Solar Observatory, New Jersey Institute of
Technology, Big Bear City, CA 92314, USA}
\affil{$^2$ \it Kiepenheuer-Institut f{\" u}r Sonnenphysik, Sch{\"o}neckstrasse
6, D-79104 Freiburg, Germany}

\begin{abstract}
\small
We observe that intergranular jets, originating in the intergranular
space surrounding individual granules, tend to be associated with granular
fragmentation, in particular, with the formation and evolution of a bright
granular lane (BGL) within individual granules. The BGLs have recently been
identified as vortex tubes by Steiner et al. We further discover the
development of a well-defined bright grain located between the BGL and the dark
intergranular lane to which it is connected. Signatures of a BGL may reach
the lower chromosphere and can be detected in off-band \ha images. Simulations
also indicate that vortex tubes are frequently associated with small-scale
magnetic fields. We speculate that the intergranular jets detected in the NST
data may result from the interaction between the turbulent small-scale fields
associated with the vortex tube and the larger-scale fields existing in the
intergranular lanes. The intergranular jets are much smaller and weaker than all
previously known jet-like events. At the same time, they appear much more
numerous than the larger events, leading us to the speculation that the total
energy release and mass transport by these tiny events may not be negligible in
the energy and mass-flux balance near the temperature minimum atop the
photosphere. The study is based on the photospheric TiO broadband (1.0~nm)
filter data acquired with the 1.6~m New Solar Telescope (NST) operating at the
Big Bear Solar Observatory. The data set also includes NST off-band \ha images
collected through a Zeiss Lyot filter with a passband of 0.025~nm.
\bigskip
\end{abstract}

\section{Introduction}

The New Solar Telescope \citep[NST][]{goode_nst_2010} of Big Bear Solar
Observatory enabled us to resolve previously unknown, even finer chromospheric
features in blue shifted off-band \ha images \citep{goode_apjl_2010}. These
absorbtion features, arising from the dark intergranular lanes, are
granular in extent, short-lived, and show jet-like upflows (Doppler shifted to
the blue). Their presence in the far blue shifted images indicates fast plasma
upflows. Their origin seems neither to be unequivocally tied to concentrations
of photospheric bright points nor predominantly associated with the vertex
formed by three (or more) granules. A visual inspection of NST data revealed
that they frequently originate from an intergranular lane separating two
granules. This is also the location where magnetic fields occur
\citep[e.g.,][]{berger&title2001,ishikawa_2008}. The projection of these
jet-like features onto the image plane may reach of typical length of 1~Mm,
while maintaining their width at 0.2~Mm during a typical lifetime of 30-60s
(some as long as 4~min).

Jets and associated chromospheric dynamics may play an important role in
chromospheric and coronal heating. Observations of active regions offered some
explanations for upflows observed in UV, such as coronal reconnection
\citep{Brooks_2008, harra_2008, baker_2009}, emerging flux \citep{Harra_2010},
as well as intermittency at the edges of active regions \citep{he_2010}, and
active region expansion \citep{murray_2010}. In quiet Sun areas and in coronal
holes, the situation is less well-understood, mainly because of limited spatial
resolution. Difficulties also arise from the fact that the chromosphere is a
quite complex layer with the complexity arising mainly from the temperature and
density distribution.

In this study, we demonstrate that the small-scale intergranular
jet-like events are clearly associated with the vortex tubes, predicted by
\citet{Steiner_2010}. The study is based on the NST photospheric (TiO) and
chromospheric (\ha) images and the comparison with model predictions.

\section{Observations}

A small coronal hole (CH) located at N03E12 was observed using the NST
\citep{goode_nst_2010, goode_apjl_2010} with adaptive optics on August 31, 2010
from 17:40 to 18:16 UT. This CH first appeared on the solar disk on August 30
and then it slowly expanded and gradually migrated in a southward direction.
According to an SDO/AIA 192\AA~ image taken at 17:46:56 UT, the CH appeared to
be magnetically connected to the neighboring NOAA AR 11101 by a set of dark
loops.

Two distinct NST data sets were acquired simultaneously with a broad-band filter
imager (BFI) using a TiO 1.0~nm filter centered at 705.7~nm and a narrow-band
filter imager (NFI) utilizing an H$\alpha$ Zeiss Lyot filter with a passband of
0.025~nm. The pixel sizes for the NFI and BFI data were 0.0375'' and 0.075'',
respectively, and the field of view of both instruments was 77''.

All time series were corrected for dark current and flat field, and then speckle
reconstructed employing the KISIP speckle reconstruction code
\citep{kisip_code}. Each reconstructed image was derived from the 70 best 1~ms
exposures (20~ms in the case of H$\alpha$ Zeiss data), acquired in quick
succession within a 7~s burst.

The resulting TiO data have a cadence of 10~s. During the observations, the Lyot
filter was switching between three wavelengths. In this observing mode, 4
consecutive images were taken at H$\alpha$-0.1~nm, and one image was taken at
H$\alpha$ line center and H$\alpha$+0.07~nm, respectively. Therefore, the groups
of 4 H$\alpha$-0.1~nm images follow each other with a 30~s interval, while the
cadence within a group is 10~s. The resulting H$\alpha$ line center and
H$\alpha$+0.07~nm images are available with a 65~sec cadence. All images in the
data sets were rigidly co-aligned using cross-correlation and then de-stretched
to remove distortions due to residual seeing effects of the Earth's atmosphere.
The final time series contained 183 TiO images and 103 H$\alpha$-0.1~nm images
and the FOV was 34''.

\section{Results}

In order to link photospheric dynamics to chromospheric activity, TiO and \ha
data sets were accurately co-aligned and combined images were produced, where
TiO granulation is overlayed with H$\alpha$-0.1~nm jet-like features.

We used this composite data set to search for sources of small-scale
intergranular jet-like events \citep{goode_apjl_2010}. Each intergranular event
was identified as follows: 1) it should be a narrow, elongated structure, 2) it
should be present in several consecutive images; 3) one end of the streak should
originate in an intergranular lane; 4) the opposite end should be evolving
(i.e., extending, shrinking, showing variations in darkening); we thus found
that intergranular events are very often associated with rapidly evolving
granules, more precisely, with instances in which granule fragments are pushed
into the intergranular lane. A more detailed analysis of granular evolution led
us to conclude that intergranular events are most often associated with what was
identified by \cite{Steiner_2010} as a vortex tube. At the same
time, not every vortex tube event was accompanied by a chromospheric jet.

\begin{figure}[t]
\centerline{\epsfxsize=3.25truein  \epsffile{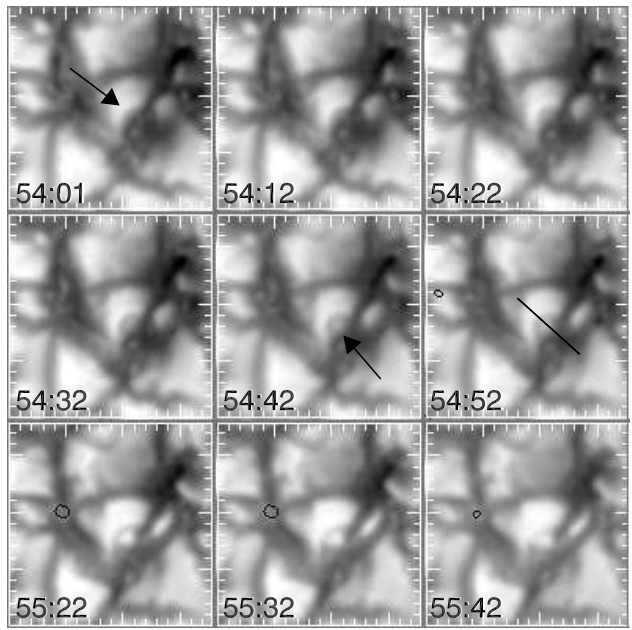}}
\caption{\small\sl Sequence of TiO images showing the development of a bright granular
lane (BGL, arrow in 54:01 frame). The arrow in the first panel indicates a BGL, which
slowly shifts toward the center of the granule as the lane evolves. The arrow in
the 54:42 frame indicates a bright grain that usually accompanies the
development of the BGL. The line segment in the 54:52 frame
marks the location of the intensity profiles plotted in Figure \ref{vt_prof}.
The time stamp in each frame shows the minute and second of image acquisition,
starting from 17:00:00~UT. Short tick marks separate 0.25~Mm spatial intervals.}
\label{vt}
\end{figure}

\begin{figure}[!th]
\centerline{\epsfxsize=3.25truein  \epsffile{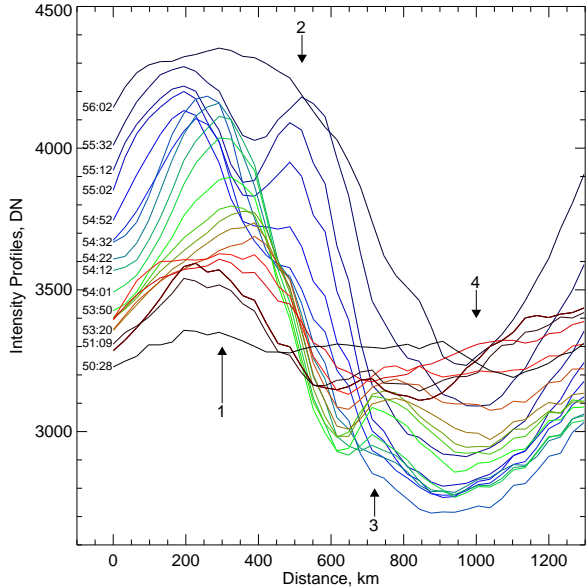}}
\caption{\small\sl Intensity profiles of a developing bright granular lane (BGL). The
profiles were measured along the line segment shown in the 54:52 frame of Figure
\ref{vt}. The origin of the $x$-axis $(x=0)$ corresponds to the upper left end
of the line segment. Time stamps (MM:SS) in the left margin of each profile
indicate the acquisition time of the corresponding image. Arrow 1 marks the
initial position of the shifting BGL, arrow 2 points to the intensifying bright
grain, arrows 3 and 4 mark the initial and late position of the dark
intergranular lane. The group of black over red to orange profiles indicate the
BGL at the early stage of the development. The green to cyan profiles refer to
times when the BGL was most dynamic, while the blue to dark blue profiles
show the late phase of development, marked by a stagnating BGL and intensely
growing bright grain(s). The colors are gradually changing to reflect the
gradual development of the event and smooth transition from one phase to
another.}
\label{vt_prof}
\end{figure}

\subsection{Development and Evolution of Bright Granular Lanes}

Figure \ref{vt} shows the evolution of a bright granular lane (BGL), which was
associated with very weak chromospheric activity. The arrow in the 54:01 panel
points to a BGL, which usually indicates the development
of a vortex tube. The BGL first forms near the edge of a granule (indicated in
the 54:01 frame), and then it is displaced toward the granule's center and one
or several bright grains appear in its place (indicated in the 54:42 frame). The
estimated speed of the BGL displacement may reach up 2~km s$^{-1}$.
\cite{Steiner_2010} proposed that the BGL is the most apparent signature of a
horizontally oriented vortex tube, with up-flows in the BGL and downflows
in the intergranular space.

Figure \ref{vt_prof} shows the evolution of intensity profiles along the line
segment shown in the 54:52 panel of Figure \ref{vt}. The profiles detail all
stages of the vortex tube's development. The red profiles reflect an early stage
(17:50 -- 17:54~UT) of development (arrow 1), during which the BGL had a low,
broad profile. Between 17:54 and 17:55~UT (green and dark blue profiles) one
side of the BGL (at $x=350$) rapidly intensified and the peak of the bright
structure was shifting toward the center of the granule (i.e., toward $x=0$~km).
After 17:54:30~UT, we observe the growth of a bright grain (indicated by arrow
2, see also the 54:42 frame in Figure \ref{vt}). By 17:55:30~UT, the intensity
of the grain nearly matched that of the BGL, and later those two structures
formed one broad bright patch at the edge of the host granule. Since neither the
BGL nor the grain showed any sense of displacement toward each other, we
speculate that the apparent merging of these two intensity structures could be
due to their horizontal expansion. Finally, after 17:56~UT the bright patch
began to fade and several minutes later the corresponding intensity profile (not
shown in the figure) was close to the pre-event profile at 17:50~UT.

The observed time profiles of the BGL are in a very good agreement with the
synthesized intensity profile presented by \cite{Steiner_2010}. In particular,
the 17:54:52 profile, where the bright grain is seen as a plateau at
$(x,y)=\{470, 3720\}$, is the best match to the snapshot shown in Fig. 5 of
Steiner et al. (2010). According to the model data, the plateau normally
develops into a structure reminiscent of the grain described here. Also, both
the observed size of the vortex tube, measured from the BGL to the intergranular
lane (0.8~Mm) and the distance between the BGL and the grain (0.3~Mm) closely
match the model data.

\begin{figure*}[!th]
\centerline{\epsfxsize=7.truein  \epsffile{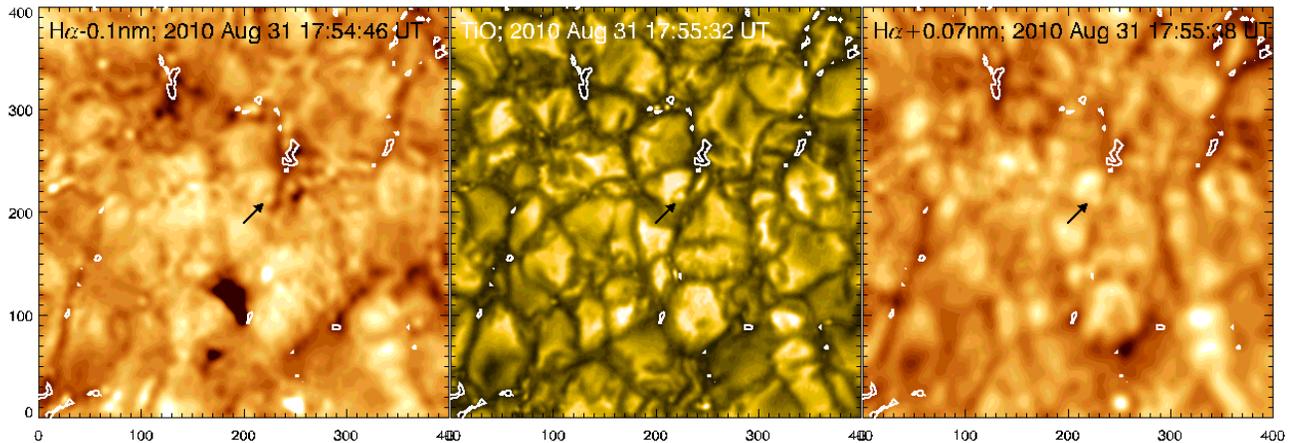}}
\caption{\small\sl The same bright granular lane as in Figures \ref{vt} and
\ref{vt_prof}
seen at three different wavelengths: \ha-0.1~nm (left), TiO band (middle) and
\ha+0.07~nm (right). The arrow in each panel points to the same bright grain.
The faint white contours shown in each panel are to facilitate image comparison
and they encircle the darkest fragments of the TiO image. The axes are image
pixels and the smallest tick marks indicate 10 pixels (0''.375 or 0.276~Mm).}
\label{tio_ha_vt}
\end{figure*}

Figure \ref{tio_ha_vt} shows three images of the above event as seen in \ha and
TiO spectral lines. The left panel is a \ha-0.1~nm image, while the right panel
shows a \ha+0.07~nm image taken nearly 1~min later. The middle panel is a TiO
image at 17:55:32U~UT. The faint white contours are the same in all three images and
they are intended to facilitate the comparison between all three panels. The
contours encircle the darkest TiO features, and they also show that the
co-aliment between the TiO and \ha data sets is not worse than 3 pixels or
0''.11. Note that the short ticks separate 10 pixel intervals. The field of view
of these images is 15''$\times$15'' or 11$\times$11~Mm. We note that both the
BGL and the grain can be identified in all three images, which means that the
vortex-tube signatures are present in the lower chromosphere. Also, this event was
associated with a faint chromospheric darkening in the blue H$\alpha - 0.1$~nm,
image (left panel), not present in the red shifted, H$\alpha + 0.07$~nm, image
(right panel), which indicates that the absorption features accompanying
vortex-tube events are chromospheric upflows.

The event shown in Figures \ref{vt} and \ref{tio_ha_vt} can be characterized as
an exemplary event, meaning that the BGL has a nearly ideal semi-circular shape.
The majority of observed BGLs are less ideal and they may appear as a nearly
straight line with (sometimes) jagged edges, especially in cases when two or
three bright grains are present.

\subsection{Chromospheric Activity Associated with Bright Granular Lanes}

\begin{figure}[h]
\centerline{\epsfxsize=3.25truein  \epsffile{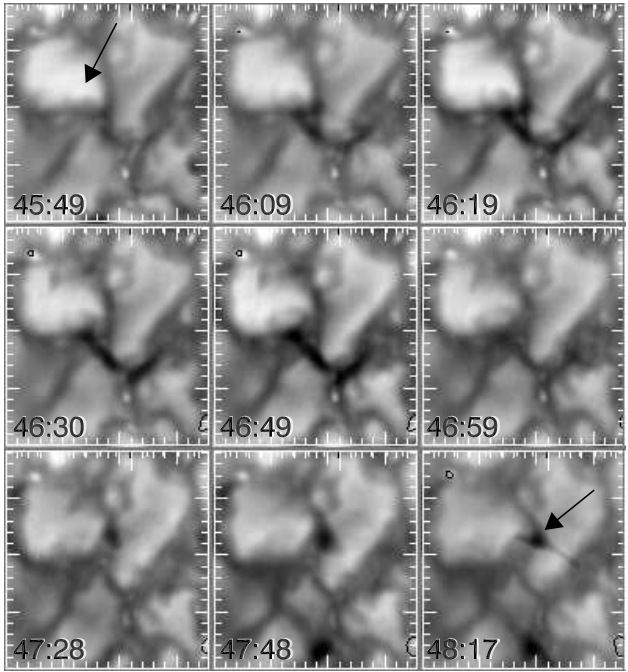}}
\caption{\small\sl A bright granular lane (BGL) associated with chromospheric
activity.
This is a sequence of composite images made by combining TiO granulation data
and dark \ha-0.1~nm absorption features. The arrow in the 45:49 frame points
toward the BGL. The onset of chromospheric jet-like activity begins
shortly after a bright grain develops (frames 46:09-46:59). The arrow in the
48:17 frame indicates an inverted ``Y'' jet occurring at approximately the same
position as the BGL. The time stamp in each frame shows the minute and
second of image acquisition starting from 17:00:00~UT. Short tick marks separate
0.25~Mm spatial intervals.}
\label{ev1}
\end{figure}

\begin{figure}[!th]
\centerline{\epsfxsize=3.25truein  \epsffile{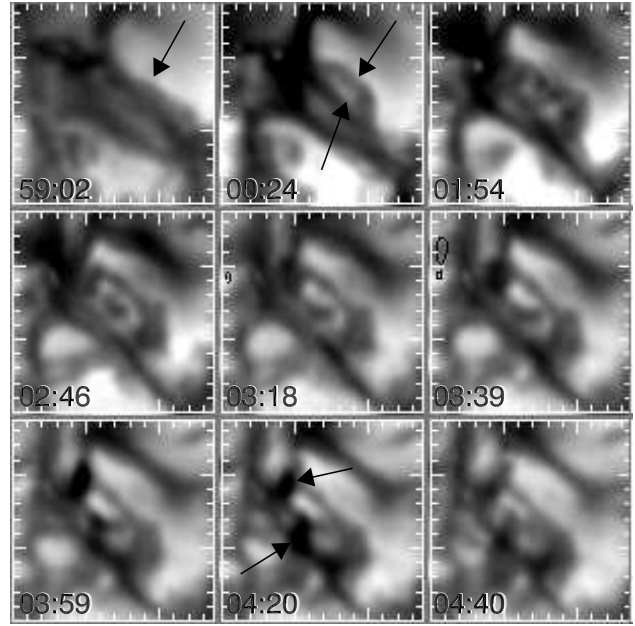}}
\caption{\small\sl Sequence of composite images showing the development of an
elongated
bright granular lane (BGL). The downward pointing arrows in the 59:02 and 00:24
frames indicate the BGL, while the upward pointing arrow marks the elongated
bright grain. Two small and short-lived absorption features appear on either
side of the grain between 18:03:39 and 18:04:40~UT. The time stamp in each frame
shows the minute and second of image acquisition starting from 17:00:00~UT.
Short tick marks separate 0.25~Mm spatial intervals, and the contour shapes in
frames 03:18 - 03:39 show bright \ha-0.1~nm patches.}
\label{ev2}
\end{figure}

At the peak of the BGL development, one or two jet-like darkenings,
co-spatial with the BGL, may appear in the blue-shifted H$\alpha$ images. These
absorption features seem to originate in the intergranular lane adjacent to the
bright grain. Figures \ref{ev1}-\ref{ev2} show three examples of chromopheric
activity (i)-(iii) associated with the formation of a BGL.

The first event began at 17:45:59~UT and lasted for about 100~s until
17:47:08~UT (Figure \ref{ev1}). The arrow in the 45:49 panel indicates the
location of a BGL, which assumed a crescent-like shape and moved toward the
center of a granule revealing a bright grain. A single strand dark
jet-like feature (i) began to develop at the same time and its intensity peaked at
about 17:46:49~UT, while the 47:28 frame shows that chromospheric feature gone.
The BGL features had faded out by about the same time. The two bottom rows of
Figure \ref{ev1} show a second small-scale event immediately following the first
one. Frame 48:17 clearly shows that the newly developed feature is reminiscent
of an inverted ``Y'' jet reported by \cite{shibata2007}. This inverted ``Y''
feature (ii) had one footpoint co-spatial with the fading BGL, while the
other one appeared to be rooted in the adjacent intergranular lane.

Figure \ref{ev2} is an example of an extended BGL, where its shape is oblong
rather than semi-circular. Similarly to other events, this BGL lane
(arrows in the 59:02 and 00:24 frames) shifted to the center of the granule
giving way to a bright grain (lower arrow in the 00:24 panel), which is
elongated in the direction of the BGL along the intergranular lane. Starting at
17:03:50~UT, two small chromospheric darkenings (iii) began to
intensify on either side of the elongated grain (04:20 frame) and these
features lasted for about 50~s.

The three types of events (i)-(iii) discussed above are the most
frequently observed types. Other, although less numerous, BGL-jet events may
include 2 or 3 \ha absorption features, two simultaneous BGLs in one
host granule and well-defined grains elongated in the direction perpendicular to
the BGL and the intergranular lane.

\section{Conclusions and Discussion}

Using NST TiO and \ha observations, we discovered that small-scale
intergranular jets, first described in \cite{goode_apjl_2010}, are associated
with bright granular lanes (BGLs) developing inside photospheric granules. The
BGLs are thought to be a signature of vortex tubes and they were first
found in solar and simulation data by \citet{Steiner_2010}. We summarize our new
findings as follows: i) our conservative estimate is that more than half of a
total of 100 well identified tiny intergranular jets are co-spatial and
co-temporal with the occurrence of BGLs, although not each BGL event is
accompanied with small-scale chromospheric activity; ii) along with the BGL, a
vortex tube also develops a well-defined bright grain located between the
BGL and the dark intergranular lane; and iii) vortex-tube signatures may reach
the lower chromosphere and can be detected in off-band \ha images. The bright
grain, described here, appears to correspond to the plateau in the model
intensity profile presented in Figure 5 in \cite{Steiner_2010}.

According to the simulation data, the darkish space between the bright grain and
the BGL coincides with the axis of the vortex tube. The interpretation is that
due to low pressure and temperature, the opacity above the vortex tube is
reduced thus allowing us to peer deeper into its relatively cooler interior.
What is the bright grain then? Is it part of the vortex tube? Does this
interpretation hold when we consider the fact that the vortex tube can reach the
chromosphere? As it follows from simulations, the associated magnetic field, is
generally wrapped up in such a way, that the field is mainly aligned with the
flow, i.e., it is rather perpendicular to the vortex tube axis.
The high-speed flow above the vortex tube reaches up to the top of the
photosphere with velocities up to 8~km s$^{-1}$ and sweeps the magnetic field
in the horizontal direction to the intergranular lane.
It may be that this field collides
with the nearby intergranular field of possibly opposing polarity, which has the
potential to cause some chromospheric activity.

We do not know, however, if BGL events seen in images of granulation possess
magnetic fields strong enough to cause detectable chromospheric activity.
Polarization measurements with the baloon-borne solar telescope {\sc SUNRISE}
failed to detect
magnetic field signal associated with a vortex tube \citep{Steiner_2010}. A
brief review of published Hinode/SP data
\citep[e.g.,][]{centeno2007,lites_2008,2010ApJ...713.1310I, gomory_2010}
indicates that Hinode/SP intensity maps may have insufficient spatial resolution
to reliably discern a BGL event, so that no reliable conclusions on the
association between a BGL and the magnetic field can be made. Nevertheless,
\citet[][left panel in their Figure 1]{centeno2007} studied a flux-emergence
event associated with a particular pattern in the intensity maps that could be
interpreted as an evolving BGL. \cite{orozco} presented data for two
flux-emergence events, where increased circular polarization polarization was
co-spatial with enhanced brightness within a granule.
\cite{Zhang_granule_fragmentation} reported that granules tend to fragment when
magnetic fields emerge within them. Simulations by \cite{tortosa_andreu} seem to
confirm the latter by showing that surface temperature structures change as
field emerges. On the other hand, \cite{gomory_2010} argue that an emerging loop
leaves no detectable brightness pattern on the host granule.

\cite{stenflo_2011} underscored the possible existence of two distinct
populations of the solar magnetic fields: i) strong, or collapsed, fields
predominantly located in the intergranular lanes and manifested via photospheric
bright points and ii) weak, or uncollapsed, flux occupying both intergranular
lanes and bright granules with a weak preference for the bright granular cells.
The uncollapsed population is thought to represent weaker turbulent fields with
spatial scales too small to be fully resolved with today's state-of-the-art
instrumentation. In this case, we suggest that the intergranular jets,
associated with the development of BGLs, may be a manifestation of these weaker
turbulent fields, the bulk of which apparently remains hidden at spatial scales
below 200~km \citep{stenflo_2011}. The intergranular jets are much smaller and
weaker than all previously known jet-like events. At the same time, they appear
much more numerous than the larger events, leading us to the speculation that
the total energy released by these tiny events may not be negligible in the
total energy balance.

\small
Authors thank BBSO observers and the instrument team for their contribution
to this study. VY work was partly supported under NASA GI NNX08AJ20G and
LWS TR\&T NNG0-5GN34G grants. VA acknowledges partial support from NSF grant
ATM-0716512. PG, VA and VY are partially supported by NSF (AGS-0745744), NASA
(NNY 08BA22G). PG is partially supported by AFOSR (FA9550-09-1-0655).
OS acknowledges insightful discussions on small-scale filament formation within
the ISSI International Team lead by I.N. Kitiashvili at ISSI (International
Space Science  Institute) in Bern.

\begin{thebibliography}{21}
\expandafter\ifx\csname natexlab\endcsname\relax\def\natexlab#1{#1}\fi

\bibitem[{{Baker} {et~al.}(2009){Baker}, {van Driel-Gesztelyi}, {Mandrini},
  {D\'emoulin}, \& {Murray}}]{baker_2009}
{Baker}, D., {van Driel-Gesztelyi}, L., {Mandrini}, C.~H., {D\'emoulin}, P., \&
  {Murray}, M.~J. 2009, \apj, 705, 926

\bibitem[{{Berger}\& {Title}(2001)}]{berger&title2001}
{Berger}, T.~E., \& {Title}, A.~M. 2001, \apj, 553, 449

\bibitem[{{Brooks} {et~al.}(2008){Brooks}, {Ugarte-Urra}, \&
  {Warren}}]{Brooks_2008}
{Brooks}, D.~H., {Ugarte-Urra}, I., \& {Warren}, H.~P. 2008, \apjl, 689, L77

\bibitem[{{Centeno} {et~al.}(2007){Centeno}, {Socas-Navarro}, {Lites}, {Kubo},
  {Frank}, {Shine}, {Tarbell}, {Title}, {Ichimoto}, {Tsuneta}, {Katsukawa},
  {Suematsu}, {Shimizu}, \& {Nagata}}]{centeno2007}
{Centeno}, R., {et~al.} 2007, \apjl, 666, L137

\bibitem[{{G\"om\"ory} {et~al.}(2010){G\"om\"ory}, {Beck}, {Balthasar},
  {Ryb\'ak}, {Ku\v{c}era}, {Koza}, \& {W\"{o}hl}}]{gomory_2010}
{G\"om\"ory}, P., {Beck}, C., {Balthasar}, H., {Ryb\'ak}, J., {Ku\v{c}era}, A.,
  {Koza}, J., \& {W\"{o}hl}, H. 2010, Astron. Astrophys., 511, A14

\bibitem[{{Goode} {et~al.}(2010{\natexlab{a}}){Goode}, {Coulter}, {Gorceix},
  {Yurchyshyn}, \& {Cao}}]{goode_nst_2010}
{Goode}, P.~R., {Coulter}, R., {Gorceix}, N., {Yurchyshyn}, V., \& {Cao}, W.
  2010{\natexlab{a}}, Astronomische Nachrichten, 88, 789

\bibitem[{{Goode} {et~al.}(2010{\natexlab{b}}){Goode}, {Yurchyshyn}, {Cao},
  {Abramenko}, {Andic}, {Ahn}, \& {Chae}}]{goode_apjl_2010}
{Goode}, P.~R., {Yurchyshyn}, V., {Cao}, W., {Abramenko}, V.~I., {Andic}, A.,
  {Ahn}, K., \& {Chae}, J. 2010{\natexlab{b}}, \apjl, 714, L31

\bibitem[{{Harra} {et~al.}(2010){Harra}, {Magara}, {Hara}, {Tsuneta},
  {Okamoto}, \& {Wallace}}]{Harra_2010}
{Harra}, L.~K., {Magara}, T., {Hara}, H., {Tsuneta}, S., {Okamoto}, T.~J., \&
  {Wallace}, A.~J. 2010, \solphys, 263, 105

\bibitem[{{Harra} {et~al.}(2008){Harra}, {Sakao}, {Mandrini}, {Hara}, {Imada},
  {Young}, {van Driel-Gesztelyi}, \& {Baker}}]{harra_2008}
{Harra}, L.~K., {Sakao}, T., {Mandrini}, C.~H., {Hara}, H., {Imada}, S.,
  {Young}, P.~R., {van Driel-Gesztelyi}, L., \& {Baker}, D. 2008, \apjl, 676,
  L147

\bibitem[{{He} {et~al.}(2010){He}, {Marsch}, {Tu}, {Guo}, \& {Tian}}]{he_2010}
{He}, J., {Marsch}, E., {Tu}, C., {Guo}, L., \& {Tian}, H. 2010, Astron.
  Astrophys., 516, A14

\bibitem[{{Ishikawa} {et~al.}(2010){Ishikawa}, {Tsuneta}, \&
  {Jur\v{c}ak}}]{2010ApJ...713.1310I}
{Ishikawa}, R., {Tsuneta}, S., \& {Jur\v{c}ak}, J. 2010, \apj, 713, 1310

\bibitem[{{Ishikawa} {et~al.}(2008){Ishikawa}, {Tsuneta}, {Ichimoto}, {Isobe},
  {Katsukawa}, {Lites}, {Nagata}, {Shimizu}, {Shine}, {Suematsu}, {Tarbell}, \&
  {Title}}]{ishikawa_2008}
{Ishikawa}, R., {et~al.} 2008, Astron. Astrophys., 481, L25

\bibitem[{{Lites} {et~al.}(2008){Lites}, {Kubo}, {Socas-Navarro}, {Berger},
  {Frank}, {Shine}, {Tarbell}, {Title}, {Ichimoto}, {Katsukawa}, {Tsuneta},
  {Suematsu}, {Shimizu}, \& {Nagata}}]{lites_2008}
{Lites}, B.~W., {et~al.} 2008, \apj, 672, 1237

\bibitem[{{Murray} {et~al.}(2010){Murray}, {Baker}, {van Driel-Gesztelyi}, \&
  {Sun}}]{murray_2010}
{Murray}, M.~J., {Baker}, D., {van Driel-Gesztelyi}, L., \& {Sun}, J. 2010,
  \solphys, 261, 253

\bibitem[{{Orozco Su\'arez} {et~al.}(2008){Orozco Su\'arez}, {Bellot Rubio},
  {del Toro Iniesta}, \& {Tsuneta}}]{orozco}
{Orozco Su\'arez}, D., {Bellot Rubio}, L.~R., {del Toro Iniesta}, J.~C., \&
  {Tsuneta}, S. 2008, Astron. Astrophys., 481, L33

\bibitem[{{Shibata} {et~al.}(2007){Shibata}, {Nakamura}, {Matsumoto}, {Otsuji},
  {Okamoto}, {Nishizuka}, {Kawate}, {Watanabe}, {Nagata}, {UeNo}, {Kitai},
  {Nozawa}, {Tsuneta}, {Suematsu}, {Ichimoto}, {Shimizu}, {Katsukawa},
  {Tarbell}, {Berger}, {Lites}, {Shine}, \& {Title}}]{shibata2007}
{Shibata}, K., {et~al.} 2007, Science, 318, 1591

\bibitem[{{Steiner} {et~al.}(2010){Steiner}, {Franz}, {Bello Gonzlez},
  {Nutto}, {Rezaei}, {Martnez Pillet}, {Bonet Navarro}, {del Toro Iniesta},
  {Domingo}, {Solanki}, {Knlker}, {Schmidt}, {Barthol}, \&
  {Gandorfer}}]{Steiner_2010}
{Steiner}, O., {et~al.} 2010, \apjl, 723, L180

\bibitem[{{Stenflo}(2011)}]{stenflo_2011}
{Stenflo}, J.~O. 2011, Astron. Astrophys., 529, A42+

\bibitem[{{Tortosa-Andreu} \& {Moreno-Insertis}(2009)}]{tortosa_andreu}
{Tortosa-Andreu}, A., \& {Moreno-Insertis}, F. 2009, Astron. Astrophys., 507,
  949

\bibitem[{{W{\"o}ger} \& {von der L{\"u}he}(2007)}]{kisip_code}
{W{\"o}ger}, F., \& {von der L{\"u}he}, O. 2007, \ao, 46, 8015

\bibitem[{{Zhang} {et~al.}(2009){Zhang}, {Yang}, \& {Jin}}]
{Zhang_granule_fragmentation} {Zhang}, J., {Yang}, S., \&
{Jin}, C. 2009, Research in Astronomy and
  Astrophysics, 9, 921

\end{thebibliography}
\end{document}